\documentclass{revtex4-1}

\usepackage{hyperref,graphicx,textcomp,mathtools}

\newcommand{\mucsi}{\mbox{\textmu c-Si:H}}
\newcommand{\mucSiC}{\mbox{\textmu c-SiC:H}}
\newcommand{\asi}{\mbox{a-Si:H}}

\begin{document}

\title{Meyer-Neldel and anti-Meyer-Neldel rule in microcrystalline silicon and
  silicon carbide examined with Hall measurements}

\author{Torsten Bronger}
\affiliation{IEK-5, Research Center J\"ulich, Germany}
\author{Oleksandr Astakhov}
\affiliation{IEK-5, Research Center J\"ulich, Germany}
\author{Reinhard Carius}
\affiliation{IEK-5, Research Center J\"ulich, Germany}

\begin{abstract}
  We study the electronic transport in lightly phosphorus-doped hydrogenated
  microcrystalline silicon (\mucsi) and nominally undoped hydrogenated silicon
  carbide (\mucSiC) by temperature-dependent Hall measurements.  The material
  properties cover different crystallinities and doping concentrations.  For
  \mucsi\ samples, the carrier concentration is altered by electron bombardment
  and subsequent step-wise annealing of defects.  We describe the behavior of
  conductivity, mobility, and carrier concentration in terms of the
  Meyer-Neldel rule (MNR) and anti-MNR.  We present the first sample switching
  between them.  A theoretical examination leverages the anti-MNR to describe
  electronic room temperature properties, and it expands the statistical shift
  model.
\end{abstract}

\maketitle

\section{Introduction}

Temperature-dependent measurements of transport parameters such as carrier
concentration, mobility, diffusion length etc.\ are an important tool for
understanding transport phenomena in semiconductors.  Very often these
parameters exhibit thermally activated behavior.  For example, the
conductivity $\sigma$ can then be calculated from its activation energy
$E_a^\sigma$ and its prefactor $\sigma_0$ according to
\begin{equation}
\sigma(T) = \sigma_0\exp(-E_a^\sigma/kT).
\end{equation}
Here, $T$ denotes temperature, and $k$ the Boltzmann constant.  Additionally
for a set of measurements, there is often observed a phenomenon called
Meyer-Neldel rule (MNR) of the form
\begin{equation}
  \label{eq:mnr}
  \sigma_0 = \sigma_{00}\exp(E_a^\sigma/E_{\mathrm{MN}}),
\end{equation}
with $\sigma_{00}$ being the Meyer-Neldel prefactor and $E_{\mathrm{MN}}$ being
the Meyer-Neldel energy.  In the Arrhenius plot, this means that the
(extrapolated) temperature-dependent lines originate all at the same point
$(k/E_{\mathrm{MN}}, \sigma_{00})$.  The case $E_{\mathrm{MN}} < 0$ is called
anti-MNR.  More generally, the MNR is visualized by plotting $\ln\sigma_0$
versus $E_a^\sigma$.  Eq.~(\ref{eq:mnr}) leads to a straight line with the
slope $E_{\mathrm{MN}}$.

Although the MNR per se is only phenomenological, it has been used frequently
for material characterization.  The MNR was first described for the
conductivity~$\sigma$ in oxide semiconductors.\cite{meyer1937relation}
Since then, $\sigma$ has remained the primary quantity for MNR examinations,
however, the list of examined materials has been extended greatly.  In
particular in the material complex of \asi, \mucsi, \mbox{a-Ge:H}, and \mucSiC,
there seems to be a universal curve $\sigma_0(E_a^\sigma)$, which approximates
the behavior of all these
materials.\cite{fuhs2000bandtails,thomas2001statistical} For low
activation energies ($E_a^\sigma<200$\,meV), there is anti-MNR with
$E_{\mathrm{MN}}\approx-20$\,meV,\cite{ram2008normal} and for higher
activation energies, there is MNR with $E_{\mathrm{MN}}\approx+40$ to
$+60$\,meV\@.\cite{thomas2001statistical,ram2002meyer,ram2008normal} For
\asi, however, even strong doping can not shift the activation energies into
the anti-MNR region.\cite{thomas2001statistical} This has been achieved,
however, with TFT structures.\cite{kondo1996observation}

In contrast to \asi, anti-MNR has been observed for \mucsi\@.  While it was
formerly assumed that only heavily-doped material may exhibit anti-MNR, it was
later reported for undoped \mucsi\ and explained with low density of grain
boundaries and extended tail states.\cite{ram2008normal} But even these samples
fit into the above-mentioned universal shape of $\sigma_0(E_a^\sigma)$.

Theoretically, the MNR in disordered silicon -- in particular the dependence
$\sigma_0(E_a^\sigma)$ -- is explained well with a statistical shift of the
Fermi level with temperature.\cite{overhof1981model,overhof1983electronic}
Almost all experimental examinations refer to $\sigma$, which is much more
accessible than~$n$.  Besides, in case of \asi, determining reliable
temperature-dependent carrier concentrations is extremely difficult due to the
sign reversal of the Hall voltage.\cite{beyer1977} This is valid with the
underlying assumption that the mobility exhibits only a weak temperature
dependence.  Strictly speaking, however, the statistical shift model applies to
carrier concentration $n$ rather than $\sigma$.

Thus, in this work we examine $n$ in terms of MNR behavior for \mucsi\ and
\mucSiC\ specimens, and compare with~$\sigma$.  In order to do so, we perform
and analyze Hall measurements.  This allows to investigate the behavior of the
hall mobility~$\mu$ as well.  Moreover, we compare the \mucSiC\ results with
the \mucsi\ results.

For our experimental program, samples of different activation energies are
needed.  There are several ways to obtain such samples.  Arguably the simplest
possibility is to vary doping concentration.  However, it is difficult to
ensure that the samples do not vary also in other parameters in this case.
Alternatively, one can use a TFT structure to vary carrier
concentration\cite{kondo1996observation} (accepting an inhomogeneous~$n$
perpendicular to the transport path), or one can use step-wise annealing of
samples degraded by the Staebler-Wronski
effect.\cite{irsigler1983application}

In this study, we vary the activation energy both by variation of doping and by
step-wise annealing of defects.  For the latter, however, we do not use the
Staebler-Wronski effect, as it is too weak in \mucsi\@.  Instead, the
respective samples are degraded using electron irradiation, which has the
additional benefit of a very wide range of defect density.

Beyond describing experimental results in terms of the MNR, it is important to
improve our understanding of the underlying physics, which still is very
incomplete.  The additional data for carrier conductivity and mobility can
provide new insight into this.  The same is true for the inclusion of \mucSiC\
into the investigation.

\section{Materials and methods}

\subsection{Preparation of silicon samples}

\begin{table}
  \caption{Doping and crystallinity data for the silicon samples.  The
    ``silane concentration'' denotes the silane concentration in
    hydrogen.  The first set are undegraded \mucsi\ samples, sample A-1 is an
    undegraded \asi\ sample, and the last set are \mucsi\
    samples intended for electron bombardment.}
  \label{tab:matrix-params}
  \begin{ruledtabular}
  \begin{tabular}{@{}lrccc@{}}
    sample & doping & silane conc. & crystallinity & thickness \\
    M-1 & 5\,ppm & 2\,\% & 84\,\%  & 2.1\,\textmu m\\
    M-2 & 10\,ppm & 2\,\% & 83\,\% & 4.0\,\textmu m \\
    M-3 & 1\,ppm & 4\,\% & 71\,\%  & 3.5\,\textmu m\\
    M-4 & 10\,ppm & 4\,\% & 74\,\% & 3.0\,\textmu m\\
    M-5 & 1\,ppm & 6\,\% & 33\,\% & 4.6\,\textmu m\\
    M-6 & 10\,ppm & 6\,\% & 38\,\% & 4.2\,\textmu m \\[1ex]
    A-1 & 10\,ppm & 7\,\% & 22\,\% & 2.7\,\textmu m \\[1ex]
    Ma-15 & 15\,ppm & 4\,\% & 79\,\% & 5.0\,\textmu m\\
    Ma-150 & 150\,ppm & 3\,\% & 79\,\% & 1.7\,\textmu m\\
  \end{tabular}
  \end{ruledtabular}
\end{table}

For this study, seven \mucsi\ layers and one \asi\ layer were prepared.
Additionally, two \mucsi\ layers were prepared intended for electron
bombardment, see next section.  All of them were deposited on roughened
borosilicate glass substrates (Corning 7059) of size 4\,\texttimes\,15\,mm$^2$
using plasma-enhanced chemical vapor deposition (PECVD) at a plasma frequency
of 95~MHz and a plasma power density of $0.07~\mathrm{W/cm^2}$.  The pressure
in the chamber was 40\,Pa, and the substrate temperature 200\,\textcelsius\@.
The deposition feed gas for the undoped films was silane diluted in
hydrogen. Doping was achieved by gas admixture of PH$_3$.  The ppm values in
this work refer to the gas phase concentration of the dopant gas with respect
to silane.

Tab.~\ref{tab:matrix-params} contains the process parameters as well as the
values of crystallinity and film thickness.  The crystallinity was determined
by Raman scattering measurements at 647\,nm excitation with the
semi-quantitative estimate for the crystalline volume fraction $I_C^{RS}
=I_{520}/(I_{520}+I_{480})$.  The film thickness was estimated from the
deposition rate and the film mass.

\subsection{Variation of defect density in the same sample}

The two silicon samples Ma-15 and Ma-150 of the above mentioned were
intended for degration, i.\,e.\ generation of defects, and subsequent stepwise
annealing.

Both films were exposed to a beam of 2\,MeV electrons with a current density
5\,\textmu A$/$cm$^2$ in a liquid nitrogen flow cryostat at approximately
100\,K\@.\cite{astakhov2007spin} Electron bombardment was performed up to
a dose of $1.1\cdot10^{18}\,\mathrm{cm}^{-2}$.  Afterwards, the samples were
handled, transported, and stored in liquid nitrogen.

The samples were annealed in vacuum at 4 or 5 temperature steps between
50\,\textcelsius\ and 190\,\textcelsius\ for 30~minutes each, and measured
after each annealing step.  Exposure time to ambient during installation of the
samples into the Hall setup was typically 3--10~minutes.

\subsection{Preparation of silicon carbide samples}

\begin{table}
  \caption{Deposition parameters and thicknesses of the \mucSiC\ samples.}
  \label{tab:sic-params}
  \begin{ruledtabular}
  \begin{tabular}{@{}lccccc@{}}
    sample & process & MMS flux & substrate & time & thick- \\
    & pressure & flux &  tempera- & in min & ness \\
    & in mbar & in sccm & ture in \textcelsius & & in nm\\[1ex]
    SIC-1 & \phantom10.7 & 6 & 330 & 200 & 300\\
    SIC-2 & \phantom10.5 & 6 & 460 & 180 & 230\\
    SIC-3 & \phantom10.5 & 10 & 400 & 75 & 300\\
    SIC-4 & \phantom11.1 & 6 & 400 & 240 & 280\\
    SIC-5 & 10.0 & 6 & 400 & 180 & 170\\
  \end{tabular}
  \end{ruledtabular}
\end{table}

Five samples of \mucSiC\ were deposited using hot-wire deposition on glass
substrate.  The varied process parameters as well as the estimated thickness
are shown in Tab.~\ref{tab:sic-params}.  For all samples, the filament
temperature was 2000\,\textcelsius\ and the hydrogen flux
approx.~94~sccm.\cite{finger2009microcrystalline}

\subsection{Hall measurements}

We structured and contacted the Hall samples photolithographically in order to
minimize offset voltages caused by an asymmetric shape of the Hall bar.  The
typical Hall bar geometry was 6\,\texttimes1.9\,mm$^2$, having two pairs of
Hall contacts.

Then, we performed temperature-dependent measurements of Hall voltage and
conductivity in the range of 80--460\,K\@.  The upper limit in temperature was
chosen to be well below the substrate temperture of the respective sample
during deposition in order to avoid changes of the sample structure.  The
measurements took place in a dark, evacuated cryostat, with a magnetic flux
through the sample of 1.9\,T\@.  The base voltage was 100\,V\@.

During the measurement, the polarity of the magnetic field was switched.  For
both polarities, the voltages perpendicular to the electrical current were
measured with electrometers with very high input resistance.  Then, the
difference of the average voltages at both polarities is twice the Hall
voltage.\cite{bronger2007carrier}

\section{Results}

\begin{figure}
  \centering
  \includegraphics[scale=0.5]{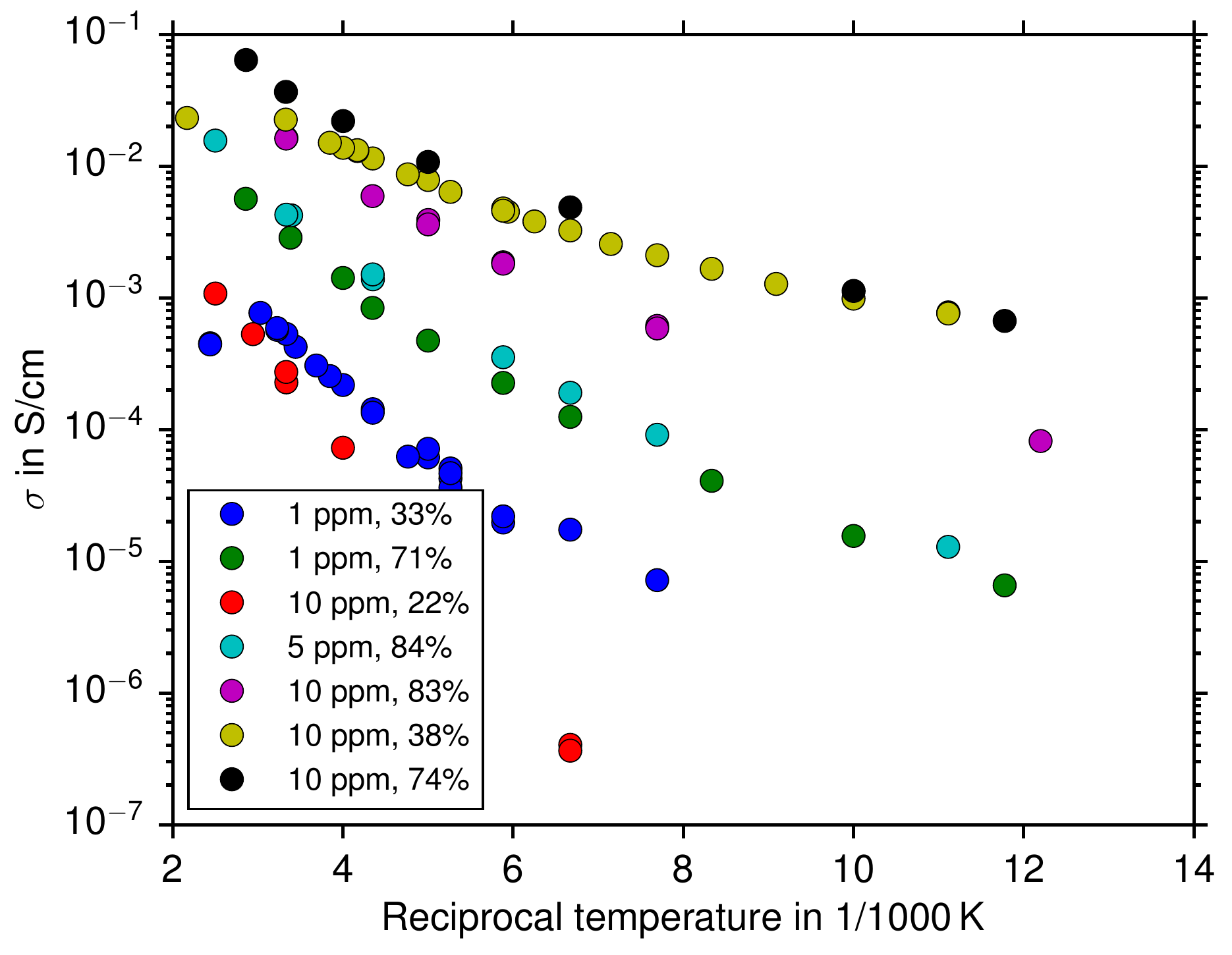}
  \caption{Temperature-dependent conductivity for the non-degraded silicon
    samples.}
  \label{fig:matrix-sigma}
\end{figure}

The temperature-dependent conductivity $\sigma$, carrier concentration~$n$, and
mobility $\mu$ of the non-degraded silicon samples as measured with the Hall
setup were already published elsewhere,\cite{bronger2007carrier}.  For
$\sigma$, they are reprinted in figure~\ref{fig:matrix-sigma}.  The sign of
carrier concentrations is negative in accordance with the n-type doping, except
for the sample with a crystallinity of only 22\,\%, which shows sign reversal,
a well-known effect for amorphous samples.\cite{beyer1977} Therefore, despite
the non-vanishing Raman crystallinity, we will refer to this sample in the
following as ``amorphous''.

All three quantities increase with increasing temperature.  Only above room
temperature, some samples exhibit a saturation effect or even a drop in $\mu$,
which for the non-amorphous samples is also visible in $\sigma$ and~$n$.  The
cause for it is unknown.  While a similar saturation was observed in photo
conductivity,\cite{ram2006study} the explanation given therein cannot be
applied here.

The curves in the Arrhenius plot are convex rather than straight, which is
least pronounced in~$n$.  The root cause for this curvature is still
speculative, with candidates being the statistical shift,\cite{yoon1986effect}
barrier height
distribution,\cite{werner1994origin,bronger2007,carius1997electronic,dyre1986phenomenological}
and differential mobility.\cite{carius1999thin} In this work, we assume the
first explanation, because it explains both curve shape and relative positions
of the curves with the same theoretical model.

\begin{figure}
  \centering
  \includegraphics[scale=0.5]{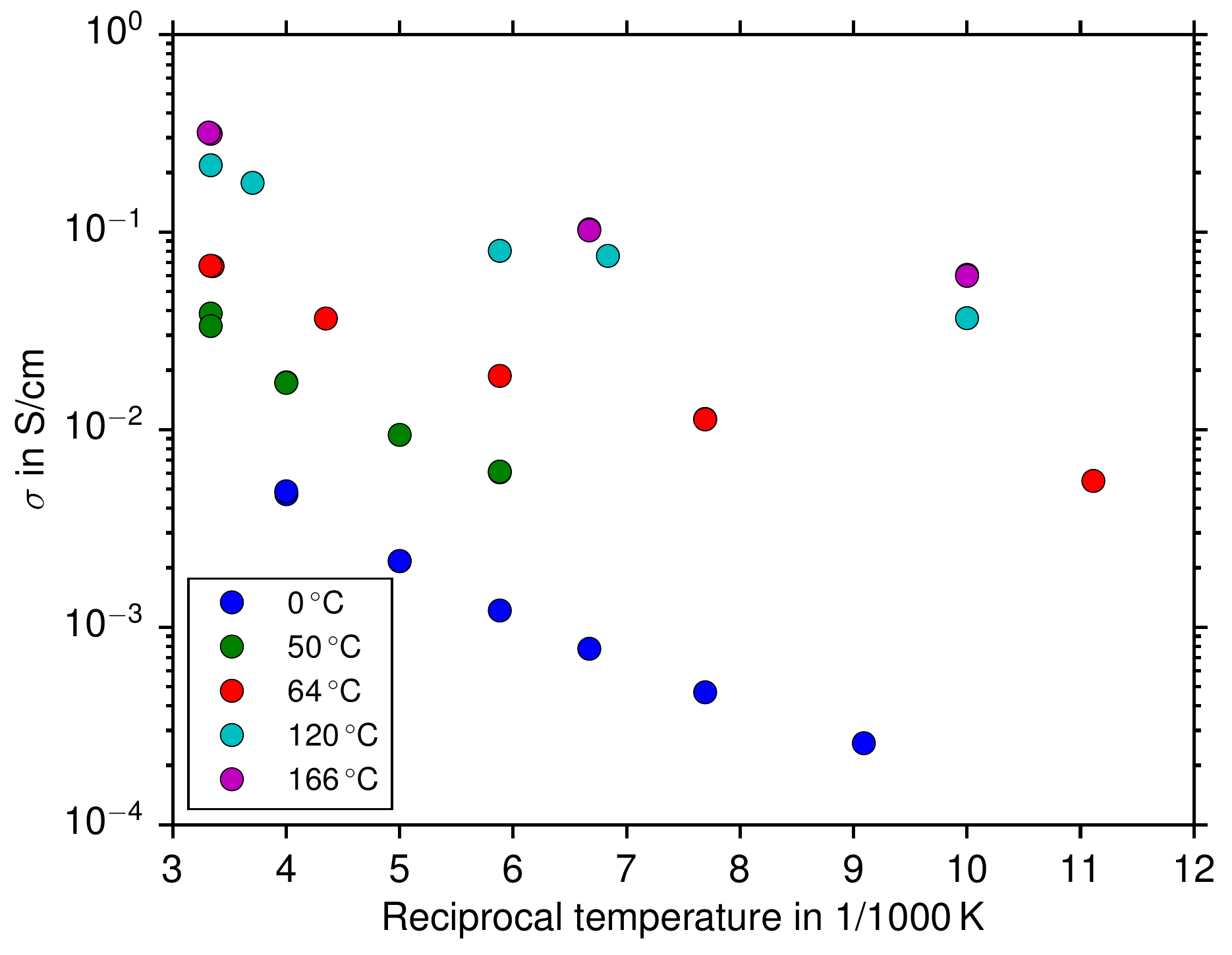}
  \caption{Temperature-dependent conductivity for the degraded sample Ma-150,
    for different annealing steps.}
  \label{fig:06B-054-sigma}
\end{figure}

\begin{figure}
  \centering
  \includegraphics[scale=0.5]{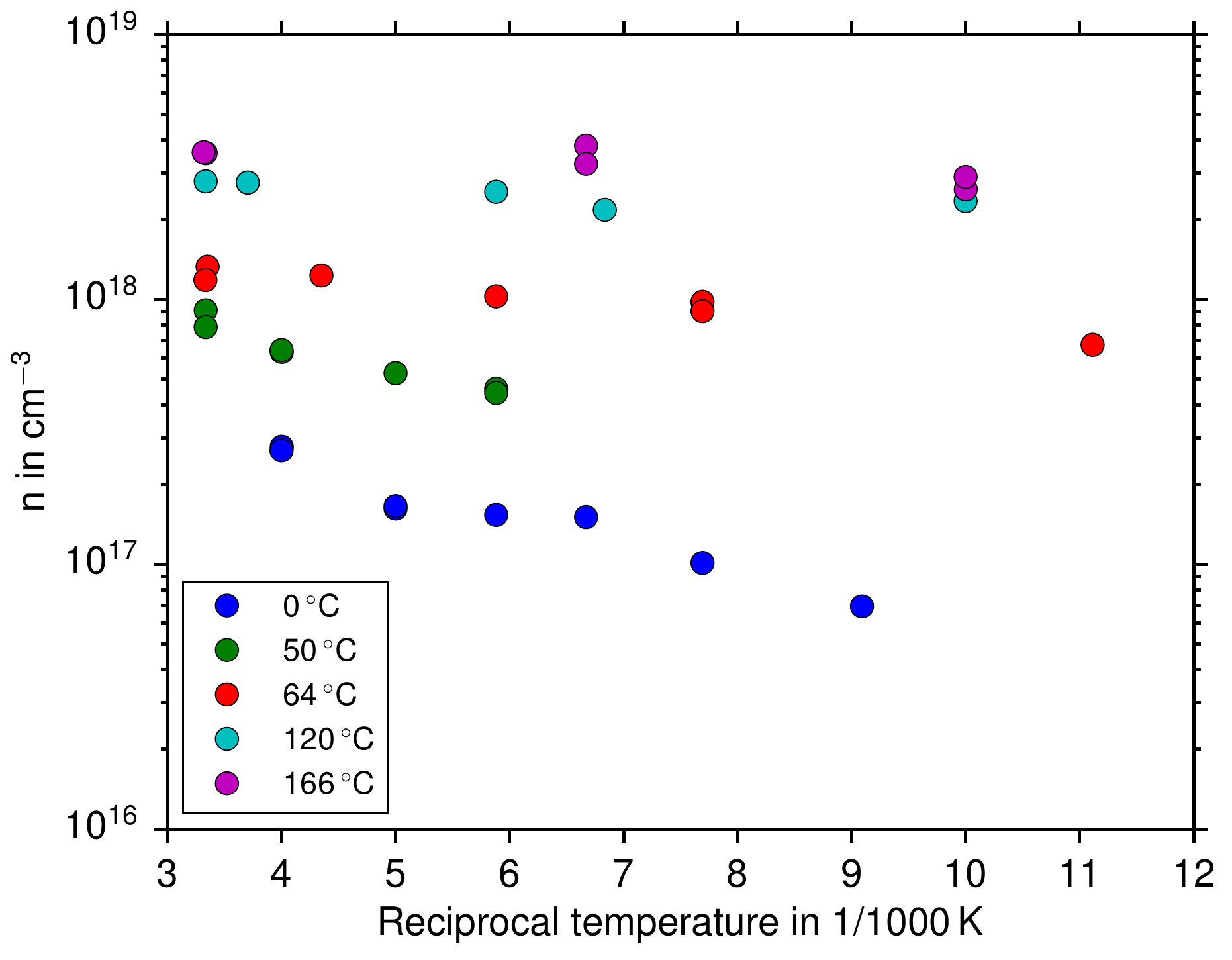}
  \caption{Temperature-dependent carrier concentration for the degraded sample
    Ma-150, for different annealing steps.}
  \label{fig:06B-054-n}
\end{figure}

\begin{figure}
  \centering
  \includegraphics[scale=0.5]{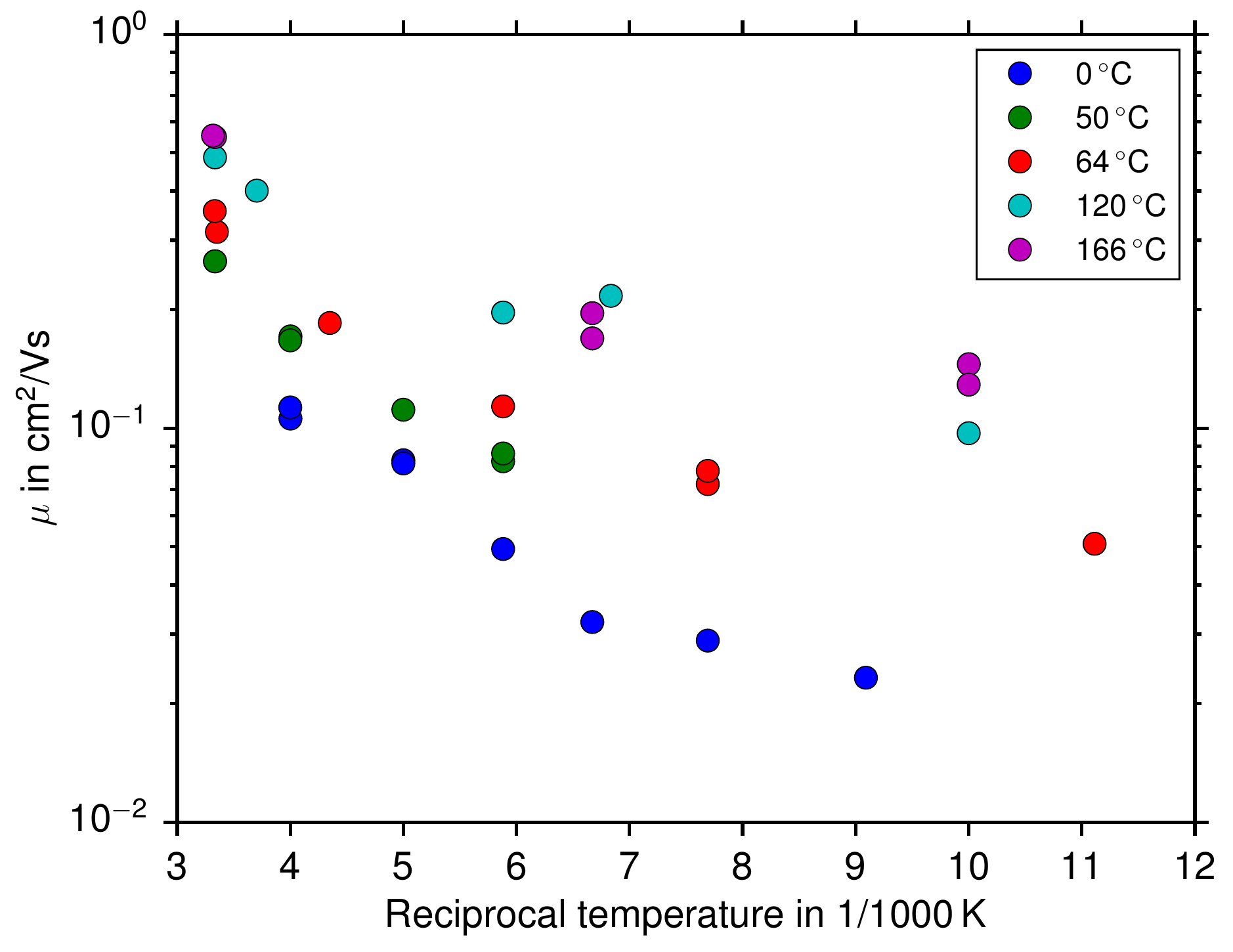}
  \caption{Temperature-dependent mobility for the degraded sample Ma-150, for
    different annealing steps.}
  \label{fig:06B-054-mu}
\end{figure}

Figures~\ref{fig:06B-054-sigma}, \ref{fig:06B-054-n}, and \ref{fig:06B-054-mu}
show the Hall results for the different annealing steps of the degraded silicon
sample Ma-150.  The curves share the important features with those of the
undegraded silicon samples, in particular the decrease to lower temperatures
and the slightly convex deviation from the Arrhenius curve.  The data points
for $\sigma$ are less scattered than those for $n$ and~$\mu$.  This visualizes
the previously made assertion that $\sigma$ is the most accessible quantity.
However, the noise in the Hall voltage was small enough to make the Arrhenius
curves for $n$ and $\mu$ almost as smooth.

\begin{figure}
  \centering
  \includegraphics[scale=0.5]{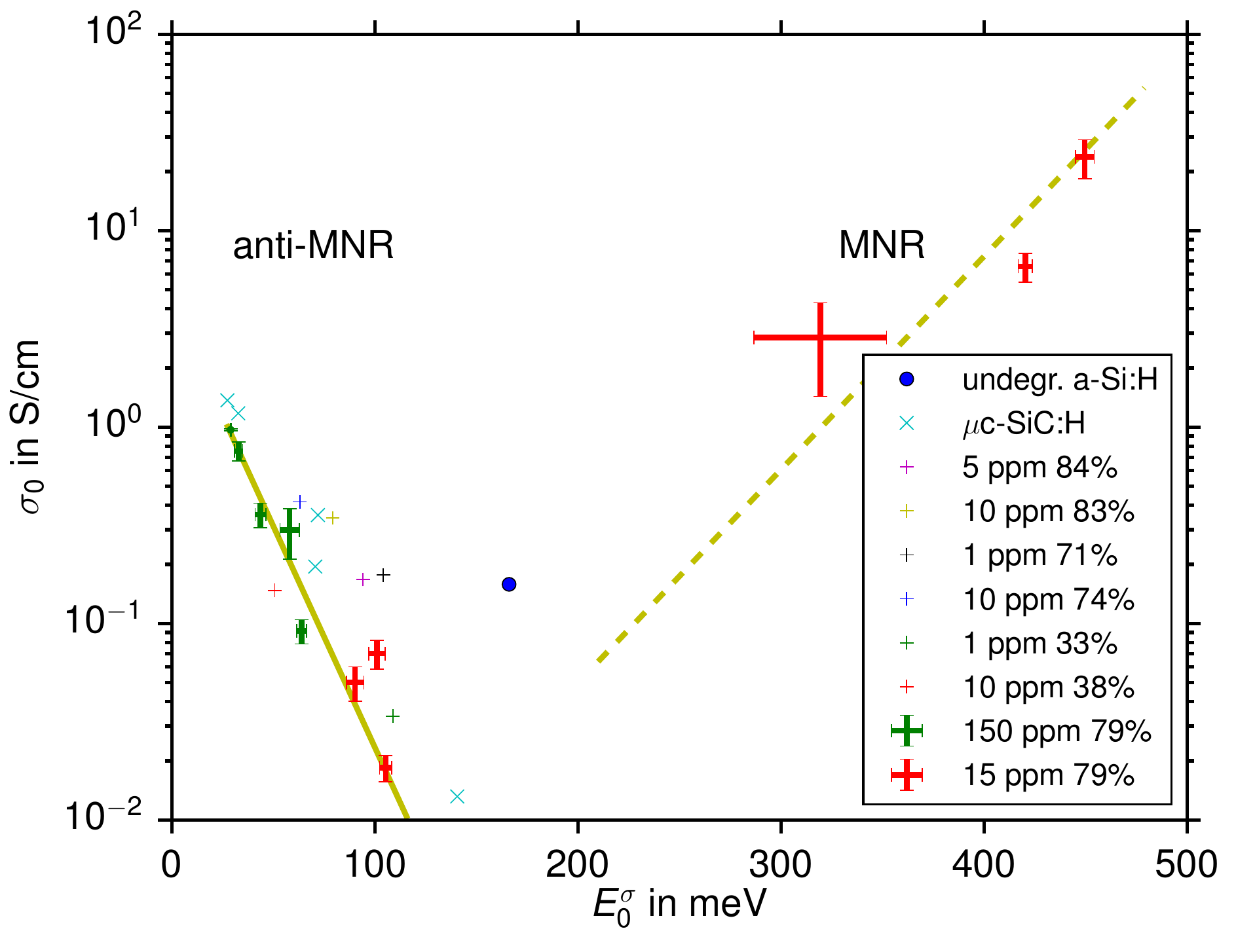}
  \caption{MNR plot of conductivity for various measurement series.  Data of
    the annealed samples (green and red) has error bars.  The yellow solid line
    is a linear fit considering only the annealed samples, see
    Tab.~\ref{tab:E_MNR}.  The dashed line is only a guide to the eye, assuming
    an MNR trend.}
  \label{fig:MNR_sigma}
\end{figure}

\begin{figure}
  \centering
  \includegraphics[scale=0.5]{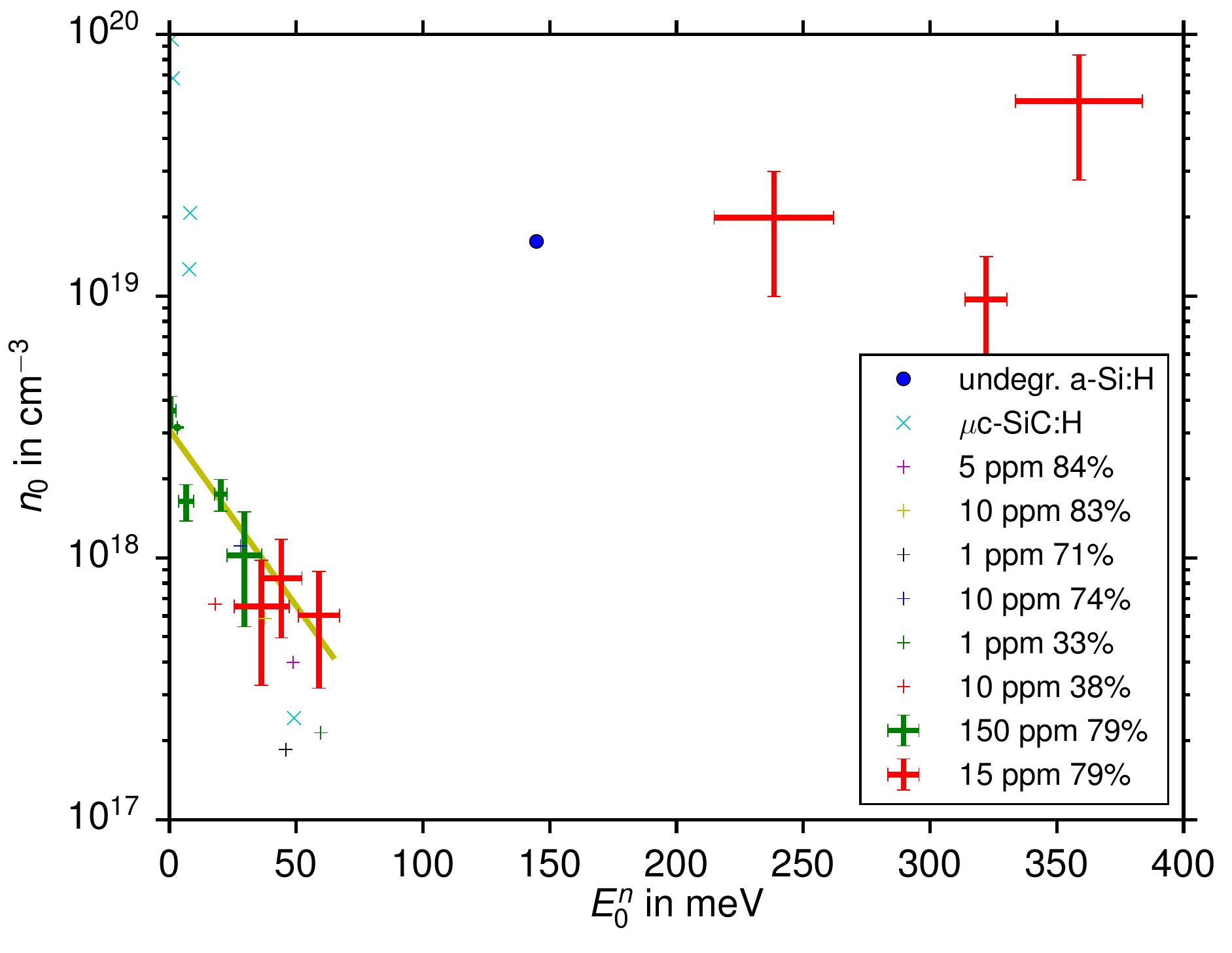}
  \caption{MNR plot of carrier concentration for various measurement series.
    The yellow solid line is a linear fit considering only the annealed
    samples, see Tab.~\ref{tab:E_MNR}.  Data of the annealed samples (green and
    red) has error bars.}
  \label{fig:MNR_n}
\end{figure}

\begin{figure}
  \centering
  \includegraphics[scale=0.5]{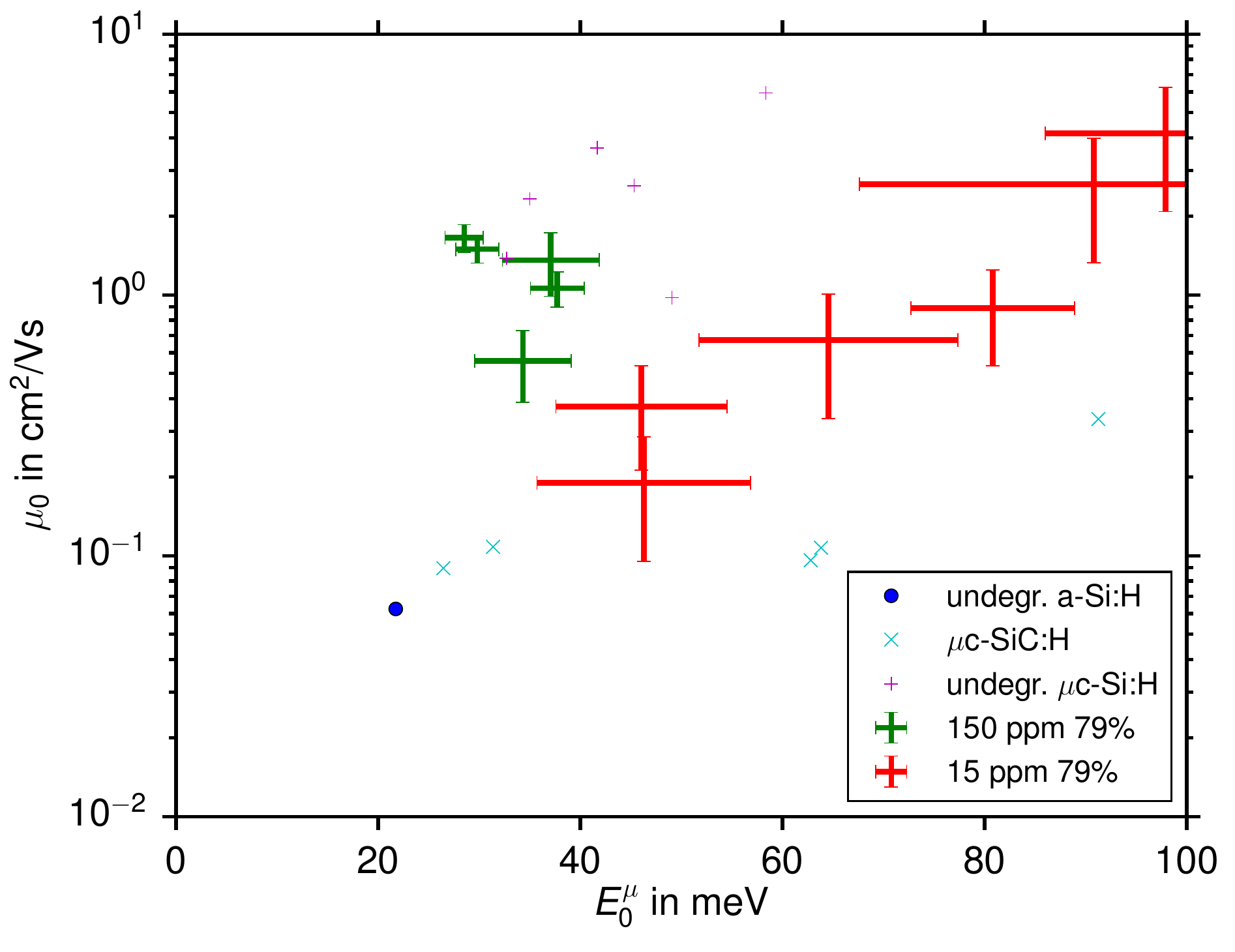}
  \caption{MNR plot of mobility for various measurement series.  Data of the
    annealed samples (green and red) has error bars.}
  \label{fig:MNR_mu}
\end{figure}

The figures~\ref{fig:MNR_sigma}, \ref{fig:MNR_n}, and \ref{fig:MNR_mu} depict
the MNR plots of $\sigma$, $n$, and $\mu$ for all samples presented here
(tables~\ref{tab:matrix-params} and~\ref{tab:sic-params}).  Every data point in
an MNR plot represents an Arrhenius curve, using its slope and $y$~intercept as
coordinates.  Note that because the original curves are generally not straight,
an MNR plot refers to a certain temperature, in this case, room temperature.  A
linear slope in the MNR plot represents the existence of a Meyer-Neldel rule.
The inverse of that slope is then the Meyer-Neldel energy $E_{\mathrm{MN}}$.
The error bars in the plots are derived from the uncertainties in slope and
intersection of the straight line fit, stemming from the scattering of the data
points of the Arrhenius curves.

\begin{figure}
  \centering
  \includegraphics[scale=0.5]{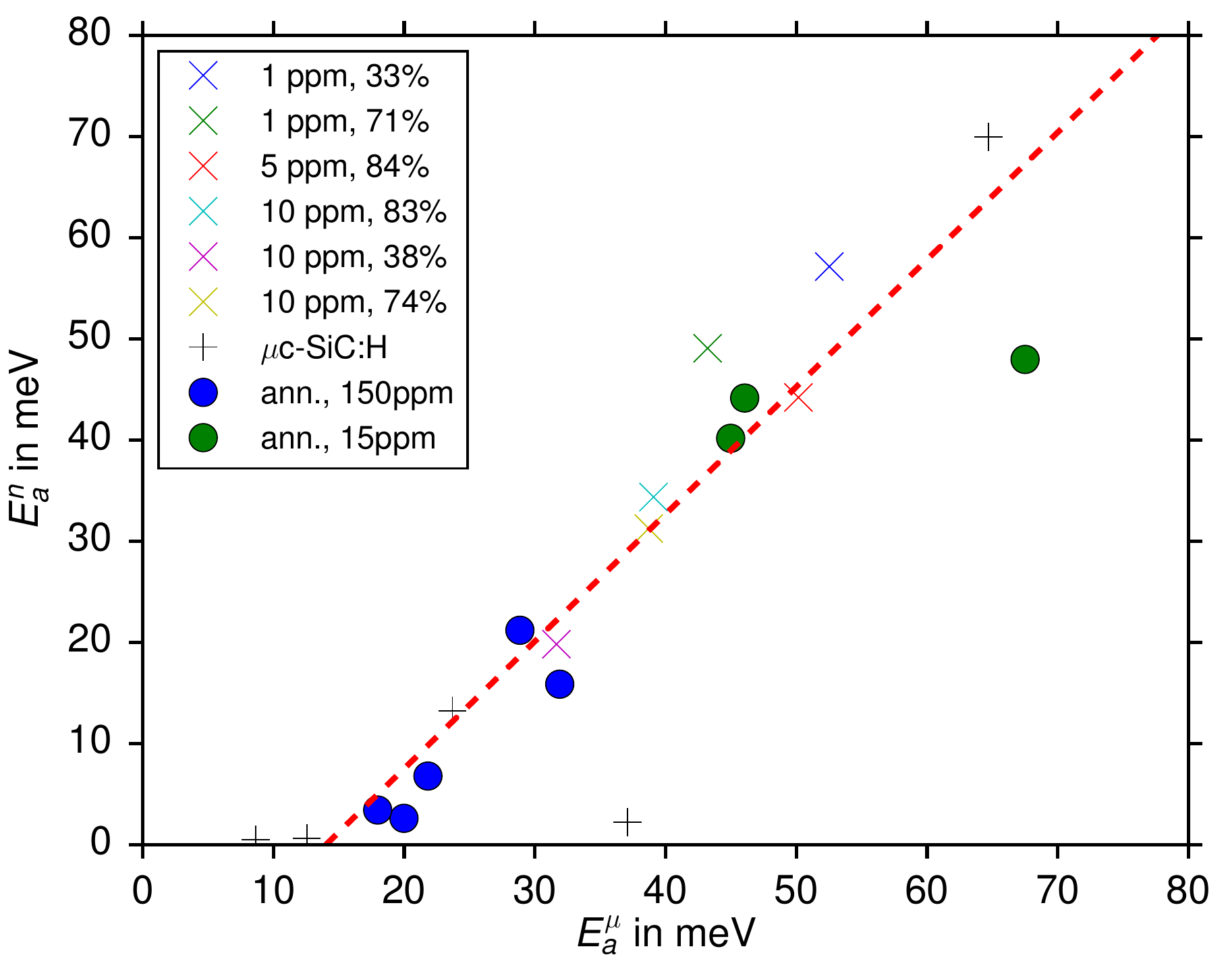}
  \caption{Activation energy of carrier concentration versus activation energy
    of mobility of various measurement series.  Only samples from the anti-MNR
    regime were taken into consideration.}
  \label{fig:Ean-vs-Eamu}
\end{figure}

Fig.~\ref{fig:Ean-vs-Eamu} shows the dependence of the activation energies of
mobility and carrier concentration.  Note that this plot, too, contains all
samples of this study, i.\,e.\ \mucSiC, undegraded microcrystalline silicon,
undegraded amorphous silicon, and annealing steps of degraded silicon.
Interestingly, most data points are positioned on the blue straight line with a
slope of 1.25 and an offset of $-18$\,meV\@.  It is important to see that only
the anti-MNR domain is covered by the ordinate.  The MNR samples have much
higher activation energies and do not follow this trend at all.

\section{Discussion}

\subsection{MNR of conductivity}

As Fig.~\ref{fig:MNR_sigma} shows, the behavior of the conductivity of all
samples can be very well described in terms of the normal Meyer-Neldel and
anti-Meyer-Neldel rules.  Normal MNR, however, is only observed for some
annealing steps of Ma-15.  The demarcation between both lies at
approx.~120\,eV, which is similar to other
reports.\cite{ram2008normal,ram2002meyer} The amorphous silicon sample
(10\,ppm/22\,\%) should also be considered being in the normal MNR regime,
although it is the only representative of its type of material, so no clear
classification is possible.

The majority of the samples and annealing steps presented in this work follow
the anti-MNR\@.  Two decades ago, this was restricted to highly doped
crystalline material.\cite{lucovsky1992transport} But general material
quality constantly has improved since then, and so even weakly doped
microcrystalline silicon can shift the Fermi level close enough to the band
edge to reach the anti-MNR domain.\cite{ram2008normal} Obviously, the
material quality of the silicon samples of this work falls in this class, too.

The \mucSiC\ samples fit in the general slope.  To our knowledge, this is the first
report of anti-MNR in \mucSiC\@.  It is astounding that even the values of
$\sigma_{00}$ and $E_{\mathrm{MN}}$ correspond to those of silicon.  This puts
even more emphasis on the apparent -- and still ununderstood -- universality of
MNR plots of disordered semiconductors which has been pointed out by other
authors in the past.\cite{fuhs2000bandtails,thomas2001statistical}

Another noteworthy result is the dichotomy of the data points of sample
15\,ppm/79\% in Fig.~\ref{fig:MNR_sigma}.  The three left-hand points lie in
the anti-MNR regime and follow the slope of anti-MNR\@.  In contrast, the three
right-hand points, despite the larger error estimates, clearly belong to the
normal MNR regime.  While such a transition was observed in \asi\ and \mucsi\
by applying different gate voltages to TFT
structures,\cite{kondo1996observation,meiling1999inverse} ours is the
first report of such a transition of a sample by changing material properties.

\subsection{MNR of carrier concentration and mobility}

The resulting MNR plots for $n$ in Fig.~\ref{fig:MNR_n} and $\mu$ in
Fig.~\ref{fig:MNR_mu} require an analysis more careful than for $\sigma$
because uncertainties are large for many of their data points.  This is due to
the small Hall voltages of these samples as well as their high electrical
resistance.  Nevertheless, it is possible to gain information from the results.

As already explained, the model of the statistical shift of the Fermi level
applies to $n$ rather than $\sigma$.  And indeed, Fig.~\ref{fig:MNR_n}
confirms corresponding MNR behavior of $n$.  In the anti-MNR regime, this
behavior is even more conformal for $n$ than for $\sigma$ comparing undegraded
and degraded silicon sample series.

As the error bars indicate, the MNR data for $\mu$ suffers from the
uncertainties more than that for~$n$.  This is suprising at first since the
main source of noise for both quantities is the Hall voltage, so the relative
error is approximately the same.  However, the activation energy interval
covered by $\sigma$ is the sum of those of $\mu$ and $n$, with $n$ taking a
much greater share.  Therefore, the uncertainties have much more impact on the
analysis of the MNR for~$\mu$.

Together with these uncertainties, the varying preparation conditions of the
undegraded silicon samples render their data points unusable for examinations
of dependences.  Thus, in order to detect a trend, it is essential to keep
sample variations small.  This is realized by considering the annealed samples
(bold red and green crosses), which reproduce the trend known from $\sigma$ and
$n$ also for~$\mu$.  However, the uncertainties do not allow a quantitative
evaluation.

\subsection{MNR in silicon carbide}

The MNR results of silicon carbide are different from those of silicon.  While
the MNR of $\sigma$ is similar to that in silicon, the anti-MNR line of~$n$ is
very steep, leading to a small $E_{\mathrm{MN}}^n$.  At the same time, $\mu$
exhibits no detectable MNR at all.

\section{Theoretical considerations}

\subsection{The significance of mobility in the MNR}
\label{sec:import-mobil-mnr}

The predictions of the statistical shift
model\cite{overhof1981model,overhof1983electronic} match our observations for
the carrier concentration $n$ well.  However, mostly the conductivity $\sigma$
is experimentally examined.  If the theory predicts MNR for $n$, but $\sigma$
is examined for MNR behavior instead, the temperature dependence of $\mu$ is of
great significance: Since
\begin{equation}
\sigma = n_{00}\mu_{00}\exp\left(\frac{E_a^n}{E_{\mathrm{MN}}^n} +
  \frac{E_a^\mu}{E_{\mathrm{MN}}^\mu}\right)
\exp\left(-\frac{E_a^n +E_a^\mu}{kT}\right),
\end{equation}
one has MNR in $\sigma$ only if
\begin{equation}
\frac{E_a^n}{E_{\mathrm{MN}}^n} +
\frac{E_a^\mu}{E_{\mathrm{MN}}^\mu} = \frac{E_a^n
+E_a^\mu}{E_{\mathrm{MN}}^\sigma} + \text{terms without $E_a$'s}.
\end{equation}
This can be satisfied only if at least one of the following conditions is met:
\begin{enumerate}
\item $E_a^\mu=0$ (i.\,e., $\mu$ is temperature-independent)
\item $E_{\mathrm{MN}}^n=E_{\mathrm{MN}}^\mu$
\item $E_a^n = a\cdot E_a^\mu+b$
\end{enumerate}
Here, $a$ and $b$ are parameters of the linear dependence.  The respective MNR
energies for $\sigma$ are:
\begin{enumerate}
\item $E_{\mathrm{MN}}^\sigma = E_{\mathrm{MN}}^n$
\item $E_{\mathrm{MN}}^\sigma = E_{\mathrm{MN}}^n=E_{\mathrm{MN}}^\mu$
\item $\displaystyle E_{\mathrm{MN}}^\sigma =
(a+1)\frac{E_{\mathrm{MN}}^nE_{\mathrm{MN}}^\mu}{aE_{\mathrm{MN}}^\mu+E_{\mathrm{MN}}^n}$
\end{enumerate}
Thus, the statistical Fermi level shift is not sufficient for the occurrence of
some form of MNR in $\sigma$.  Additionally, the mobility must fulfill at least
one of the three conditions.

Generally, examinations of the MNR in disordered semiconductors implicitly
assume~(1), i.\,e.\ a very weak temperature dependence of mobility.  This stems
from the theory of the mobility edge, above which electronic transport is
supposed to be only sub-exponentially depending on temperature: $\mu\sim
T^{-1}$~\cite{mott2012electronic}.  However, this only holds if no other
obstacles for the transport have significant impact.  There is no direct
experimental evidence for a vanishing $E_a^\mu$ in \asi, and it is certainly
not true in \mucsi\ and \mucSiC\ according to figure~\ref{fig:Ean-vs-Eamu}.

\begin{table}
  \caption{Anti-MNR energies and $y$~intercepts for $\sigma$ and $n$ for the
    annealed samples.  For $\mu$, the data uncertainties were too large for
    evaluation.}
  \label{tab:E_MNR}
  \begin{ruledtabular}
  \begin{tabular}{@{}lcl@{}}
    & $E_{\mathrm{MN}}$ (meV) & $y$ intercept \\
    $\sigma$ & $-17$ & $6.4~\mathrm{S/cm}$ \\
    $n$      & $-29$ & $3.3\cdot10^{18}~\mathrm{cm^{-3}}$ \\
  \end{tabular}
  \end{ruledtabular}
\end{table}

As shown in Tab.~\ref{tab:E_MNR}, $E_{\mathrm{MN}}^n$ exceeds
$E_{\mathrm{MN}}^\sigma$ by a factor of 1.7.  The numbers must be taken with
care due to the scattering of the measurement results, but it is safe to assert
that $E_{\mathrm{MN}}^n \ne E_{\mathrm{MN}}^\sigma$.  Thus, condition~(2) is
not met at least in the anti-MNR domain.

However, the universal proportionality of $E_a^n$ and $E_a^\mu$ as demonstrated
in Fig.~\ref{fig:Ean-vs-Eamu} does fulfill condition~(3).  Note that for this
condition, even $E_{\mathrm{MN}}^\mu=\infty$ is allowed, as seems to be the
case for the \mucSiC\ samples.  Unfortunately, to our knowledge, the dependence
$E_a^n\sim E_a^\mu$ hasn't been investigated or even explained hitherto.
(Jackson claims in \cite{jackson1988connection} that ``the conductivity in
\asi\ exhibits a MNR because the statistical shift causes the carrier density
to follow a MNR as well as the mobility'' but the given citations do not
confirm this.)  The validity of this would provide great insight into transport
in disordered semiconductors in general, and the MNR effects discussed here in
particular.

\begin{figure}
  \centering
  \includegraphics[scale=0.5]{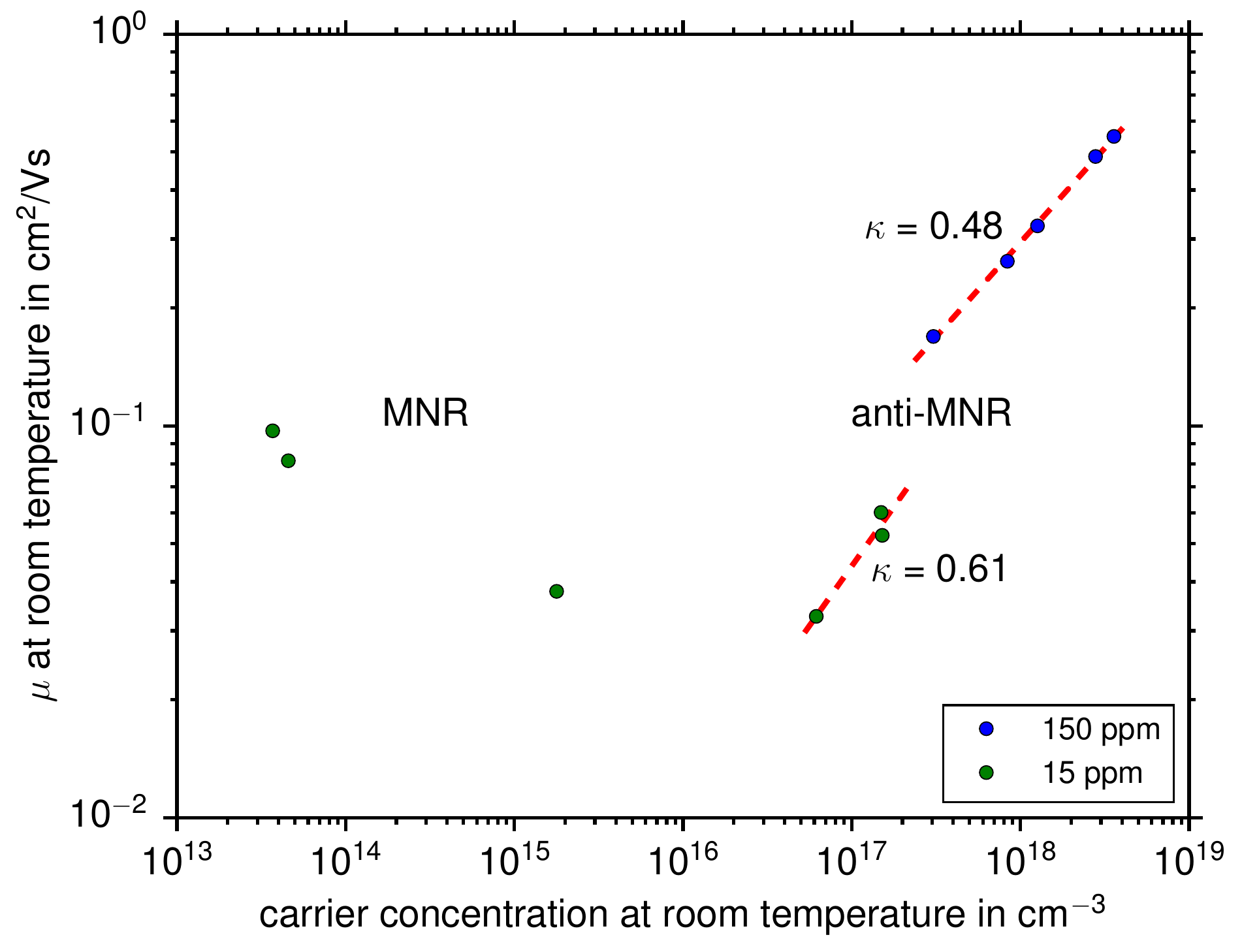}
  \caption{The mobility versus the carrier concentration of the annealed
    samples, both measured at room temperature.  The given values for $\kappa$
    represent the exponent of the poylnomial fits (printed in dashed lines) and
    refer to Eq.~\ref{eq:kappa}.}
  \label{fig:mu-versus-n}
\end{figure}

\subsection{Relation between carrier concentration and mobility at room
  temperature}

According to \cite{bronger2007carrier}, there is a correlation between the room
temperature mobility and carrier concentration $\mu_r$ and $n_r$ for n-doped
microcrystalline silicon.  The higher the doping and thus the carrier
concentration, the higher the mobility.  Due to the significance of these
properties, there is particular interest in their inter-relation.  However, no
quantitative description is given.

Fig.~\ref{fig:mu-versus-n} shows this dependence for the annealed samples used
in this work.  For 05B-054, there is a clear polynomial relation between both
quantities.  For 06B-276, there is no global trend.  However, its three data
points in the anti-MNR regime match the behaviour of 05B-054 well.  Thus,
despite the limited data available, it is plausible to assume a polynomial
relation as well.

This polynomial relation can be derived from the presence of anti-MNR and the
linear dependence of the activation energies of $n$ and $\mu$:
\begin{align}
  \mu_r(n_r) &= \mu_{00}'\left(\frac{n_r}{n_{00}}\right)^\kappa\quad\text{with}\\
  \mu_{00}'  &= \mu_{00}\exp\left(\frac{E_{\mathrm{MN}}^\mu - kT_r}{a
      E_{\mathrm{MN}}^\mu kT_r}\cdot b\right)\\
  \label{eq:kappa}
  \kappa &= \frac{E_{\mathrm{MN}}^n(E_{\mathrm{MN}}^\mu -
    kT_r)}{aE_{\mathrm{MN}}^\mu(E_{\mathrm{MN}}^n - kT_r)},
\end{align}
with $T_r$ being room temperature, $a$ and $b$ as defined in the previous
section, and $\kappa$ ranging between 0.5 and 1.5, in accordance with the large
uncertainties in $E_{\mathrm{MN}}^n$ and $E_{\mathrm{MN}}^\mu$\@.  See
appendix~\ref{sec:deriv-mu-n} for details of the calculation.

\section{Conclusions}

The electrical transport in disordered semiconductors was investigated by
evaluating Hall measurements for a critical examination of the MNR\@.  On the
one hand, it is backed by the well-established theory of the statistical shift,
and on the other hand, it allows to average many measurements, thus stabilizing
the results significantly.

The splitting of $\sigma$ into $n$ and $\mu$ for state-of-the-art, weakly
n-doped \mucsi\ samples shows that all three quantities follow an anti-MNR\@.
Moreover, \mucSiC\ samples exhibit anti-MNR for $\sigma$ and $n$ very similar
to \mucsi, in both prefactor and MNR energy.  The same could not be observed
for $\mu$, but this needs to be investigated further.  Furthermore, we
demonstrate a switch from MNR to anti-MNR in microcrystalline silicon in the
same sample by variation of its defect density, which consolidates the Fermi
level position as being critical for MNR behavior.

Theoretical examination reveals that from MNR behavior in $n$ does not
necessarily follow MNR behavior in $\sigma$, as it is necessary for using
$\sigma$ as the observable.  Instead, the temperature dependence of $\mu$ must
meet any of the conditions listed in section~\ref{sec:import-mobil-mnr}.  We
show that one of these conditions, namely the proportionality of the activation
energies of $n$ and $\mu$, is indeed fulfilled for the anti-MNR domain in both
\mucSiC\ and \mucsi, even with the same parameters.

Finally, from this proportionality, we derive a power law relation between
carrier concentration and mobility, which can be observed clearly when varying
the defect density.

Further investigations should focus on the peculiar relation of mobility and
carrier density.  So far, the behavior of the mobility has been only described
rather than understood.  The proportionality of the activation energies of $n$
and $\mu$ appears to be crucial for the understanding of the underlying physics
in the presence of anti-MNR.  Other material systems should be tested for this
proportionality, and whether their parameters are the same as for the material
systems presented here.

\appendix

\section{Derivation of \textit\textmu(\textit n) at root temperature}
\label{sec:deriv-mu-n}

Let $n_r$ and $\mu_r$ be the respective quantity measured at $T_r$.  Then, the
activation energies are difference quotients:
\begin{align}
  E_a^n &= \frac{\ln \frac{n_{00}}{n_r}}{\frac1{kT_r} - \frac1{E_{\mathrm{MN}}^n}},\\[1ex]
  E_a^\mu &= \frac{\ln \frac{\mu_{00}}{\mu_r}}{\frac1{kT_r} - \frac1{E_{\mathrm{MN}}^\mu}}.
\end{align}
Since we can connect both equations by using
\begin{equation}
E_a^n = a\cdot E_a^\mu + b
\end{equation}
(see section~\ref{sec:import-mobil-mnr}), it indeed follows
\begin{equation}
\mu_r = \mu_{00}'\left(\frac{n_r}{n_{00}}\right)^\kappa,
\end{equation}
with $\mu_{00}'$ and $\kappa$ defined as at the end of
section~\ref{sec:import-mobil-mnr}.

\bibliography{transport}

\end{document}